\documentclass[runningheads]{llncs}
\usepackage[T1]{fontenc}
\usepackage{graphicx}
\usepackage{subcaption}
\usepackage{xcolor}
\usepackage{booktabs}
\usepackage{hyperref}
\usepackage{multirow} 
\usepackage{array}
\usepackage{amssymb}

\definecolor{bram}{rgb}{1, .75, 0}
\definecolor{lucas}{rgb}{.55, .71, 0}
\definecolor{konrad}{rgb}{1, .44, .37}
\begin{document}
\title{Are Companies Taking AI Risks Seriously?\\ A Systematic Analysis of Companies' AI Risk Disclosures in SEC 10-K forms\thanks{To be published at the SoGood workshop of ECML PKDD 2025}}
\titlerunning{AI Risk Disclosures in SEC Filings}
\author{Lucas G. Uberti-Bona Marin\thanks{Equal contribution}\orcidID{0000-0001-6518-9535} \\
Bram Rijsbosch\inst{**}\orcidID{0009-0000-0803-211X}\\
Gerasimos Spanakis\orcidID{0000-0002-0799-0241}\\
Konrad Kollnig\orcidID{0000-0002-7412-8731}}
\authorrunning{Uberti-Bona Marin et al., 2025}
\institute{Law and Tech Lab, Maastricht University, the Netherlands}
\maketitle
\begin{abstract}
As Artificial Intelligence becomes increasingly central to corporate strategies, concerns over its risks are growing too. In response, regulators are pushing for greater transparency in how companies identify, report and mitigate AI-related risks. In the US, the Securities and Exchange Commission (SEC) repeatedly warned companies to provide their investors with more accurate disclosures of AI-related risks; recent enforcement and litigation against companies' misleading AI claims reinforce these warnings.
In the EU, new laws -- like the AI Act and Digital Services Act -- introduced additional rules on AI risk reporting and mitigation.
Given these developments, it is essential to examine \textit{if} and \textit{how} companies report AI-related risks to the public.

This study presents the first large-scale systematic analysis of AI risk disclosures in SEC 10-K filings, which require public companies to report material risks to their company. We analyse over 30,000 filings from more than 7,000 companies over the past five years, combining quantitative and qualitative analysis.

Our findings reveal a sharp increase in the companies that mention AI risk, up from 4\% in 2020 to over 43\% in the most recent 2024 filings. While legal and competitive AI risks are the most frequently mentioned, we also find growing attention to societal AI risks, such as cyberattacks, fraud, and technical limitations of AI systems. However, many disclosures remain generic or lack details on mitigation strategies, echoing concerns raised recently by the SEC about the quality of AI-related risk reporting.
To support future research, we publicly release a web-based tool for easily extracting and analysing keyword-based disclosures across SEC filings.
\end{abstract}
\section{Introduction}
Artificial Intelligence is rapidly transforming industries, with companies increasingly adopting AI at the core of their businesses. Yet, as AI technologies become more powerful and widely used, societal concerns are growing too \cite{bengio2024managing,center2023ai,techreport}. These growing concerns have intensified calls for stronger regulatory oversight and greater corporate accountability in how AI is developed and used \cite{bengio2024managing}. 
In response, recent legal initiatives, such as the 2024 EU AI Act and the 2022 EU Digital Services Act \cite{eurlex2024aiact,eurlex2022dsa}, have introduced new requirements for companies to identify, assess and disclose AI-related risks. In the United States, the Securities and Exchange Commission (SEC) has repeatedly cautioned companies against “AI-washing”, stressing the need for more accurate disclosures of AI risks \cite{gensler2024aiwashing,sec2024-gensler-aiwashing2,sec2024-gensler-yale3,sec2024-disclosure-rules}.
Recent SEC enforcement and private litigation against misleading AI claims reinforce these warnings \cite{alston2024-ai-enforcement,sec2025enforcementaction}.
% Research Objective
Given these developments, it is essential to examine \textit{if} and \emph{how} companies currently report on AI-related risks in legally mandated public disclosures. 

This study therefore presents the first systematic large-scale empirical analysis of AI risk disclosures in SEC 10-K filings, annually submitted by all public companies in the US. 10-K filings require firms to provide a comprehensive overview of their business, including ``information about the most significant risks that apply to the company or to its securities'' \cite{sec2024a10k}. Insufficient SEC 10-K disclosures may breach US securities law, potentially leading to litigation or even criminal charges against companies' executives \cite{sec2024a10k,alston2024-ai-enforcement}. This leads to generally high quality of the disclosures.
Companies subject to these disclosures include major tech companies (e.g. Microsoft, NVIDIA and Apple) and also large non-tech companies (e.g., Coca-Cola and Walmart). 
\begin{figure}[t]
    \centering
    \includegraphics[width=\linewidth]{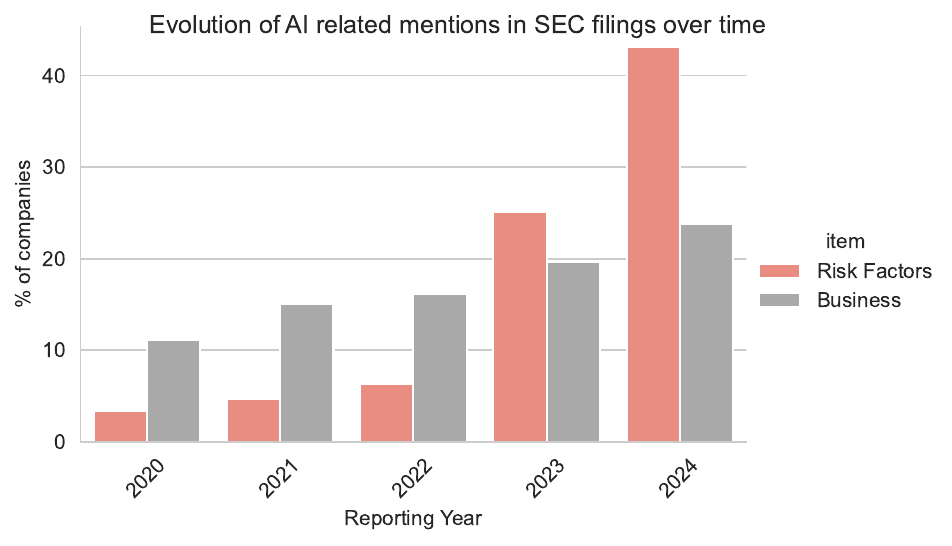}
    \caption{Percentage of companies mentioning AI in the Risks and Business sections of 10-K filings, showing how companies embrace AI and mention risks.}
    \label{fig:risks_business}
\end{figure}

Prior research on corporate risk reporting in SEC filings -- particularly in the context of climate change and cybersecurity -- has shown that companies often downplay risks until regulatory efforts force greater transparency \cite{doran2008climate,godawska2024environmental,morse2017sec,li2018sec,calderon2022changes,pengl2022security}. While several industry-led reports have noted a rise in AI-related (risk) mentions within subsets of SEC filings \cite{orrick2024aifilings,weil2023aidisclosures,deloitte2024uscmarshall,arize2024rise,caq2024aireporting}, these analyses generally lack methodological rigour and transparency, scale, and data from the most recent filings (covering reporting year 2024). 
As public concern and regulatory scrutiny over AI risks continue to grow,  a clear gap remains in our understanding of how firms currently frame and disclose AI-related risks in legally mandated disclosures.

To address this gap, we conduct a comprehensive analysis of 10-K filings from more than 7,000 public US companies over the past five years. We address two research questions: 
\begin{itemize}
    \item \textbf{RQ1 (Quantity of Disclosures):} How commonly do public US companies disclose AI use and risks in their 10-K filings, and how has this evolved over the past five years?
    \item \textbf{RQ2 (Substance of Disclosures):} What types of AI-related risks do public US companies disclose and how do these disclosures align with broader societal concerns about AI?
\end{itemize} 
%\noindent\textbf{Methodology.} To address RQ1, we conduct a large-scale keyword analysis, looking at temporal trends in AI-related (risk) mentions across our dataset (Figure \ref{fig:risks_business}). For RQ2, we conduct an in-depth qualitative analysis into the content and framing of AI risk disclosures, focusing on both the largest technology firms and a sample of randomly selected companies, hereby using a manual labelling strategy. In particular, we use the comprehensive MIT AI Risk Taxonomy \cite{slattery2024ai}, to assess which societal AI risks are being reported by companies.
%\noindent\textbf{Contributions.}
Our contributions are twofold:
\begin{enumerate}
    \item \textbf{Empirical insights and dataset of AI risks.} Our findings provide an empirical foundation for shaping discussions on how to ensure that company AI risk disclosures support meaningful transparency and accountability. We share our annotated dataset to facilitate future research.
    \item \textbf{Web tool for analysing topical disclosures in SEC filings.} We introduce a publicly available \href{https://github.com/lucas-ubm/sec_analysis}{web tool} that allows users to extract and analyse topical disclosures across SEC filings easily, supporting further research.
\end{enumerate}
%\noindent\textbf{Structure.}
The remainder of this paper is organised as follows. Section 2 provides background on SEC filings and reviews related work on risk disclosures in SEC filings. Section 3 presents the methodology and findings from the large-scale trend analysis (RQ1). Section 4 reports on our deep-dive analysis looking at what specific AI risks are being disclosed by companies (RQ2). We close by summarising our findings and discussing their implications for policy and regulation. 
\section{Background}
This section provides context on SEC 10-K filings and reviews prior work on (AI) risk disclosures analyses of 10-K filings.

\subsection{SEC 10-K filings and US securities law}
Under US securities law, all public companies must annually file 10-K reports. These are comprehensive disclosures of a company’s business, operations, risks, and financial performance \cite{sec2021howtoread}. 10-K filings typically offer more detail than annual reports and all filings are publicly available via the Edgar database \cite{sec2021howtoread,sec_edgar}.

10-K filings follow a standardised format. Two key sections include: 1) The \textit{Business} section (known as ``item 1''), which 
describes a company's main operations, products, services, and markets, and 2) the \textit{Risk Factors} section (``item 1A''), where companies are legally required to disclose the most significant risks that apply to a company or its securities \cite{sec2021howtoread}.

Filings are legally binding. False or incomplete disclosures can lead to SEC enforcement, private litigation, or even criminal charges. As such, these reports tend to contain detailed descriptions of risks, with for example an average of 13 pages of risk-related content in recent filings of Fortune 500 companies \cite{deloitte2024uscmarshall}.

The SEC plays an active role in reviewing filings and enforcing disclosure rules, with 583 enforcement actions and \$8.2 billion in financial remedies in fiscal year 2024 alone \cite{sec2024enforcement}.
In recent years, the SEC has placed growing emphasis on AI-related disclosures. Former SEC chair Gary Gensler has repeatedly warned companies against ``AI-washing'' (exaggerated or misleading claims about AI), emphasising the need for more truthful, tailored and complete disclosures of AI activities and risks \cite{gensler2024aiwashing,sec2024-gensler-aiwashing2,sec2024-gensler-yale3}. The 2024 disclosure review further outlines the SEC's expectations for 2025, which includes properly defining AI use, disclosing more company-specific AI risks, and substantiating AI claims. \cite{sec2024-disclosure-rules}. 

Several recent enforcement actions by the SEC targeting companies that made false or misleading AI-related statements have already resulted in fines  \cite{alston2024-ai-enforcement,sec2025enforcementaction}. 
At the same time, shareholder lawsuits have been filed against companies accused of failing to adequately inform about risks associated with their use of AI \cite{alston2024-ai-enforcement}. These developments highlight the growing regulatory attention and underscore the relevance of this data source. 

\subsection{Prior research on risk disclosures in SEC filings}
There is extensive prior research that has looked at how companies disclose systematic or emerging risks in their 10-K filings. Studies on climate risks, for instance, show that companies have in the past often under-reported these risks until regulatory interventions forced more robust reporting \cite{doran2008climate,godawska2024environmental}. Similar patterns have been observed in studies on cybersecurity disclosures \cite{morse2017sec,li2018sec}. In both domains, empirical research has contributed to the SEC issuing new guidance and rules to improve the quality and consistency of disclosures. 

Studies in the cybersecurity domain have shown that companies extend their risk reporting following the release of new SEC guidelines \cite{morse2017sec,li2018sec,calderon2022changes,pengl2022security}. However, there is an ongoing debate whether this also improves the quality of those disclosures \cite{pengl2022security,godawska2024environmental}. In light of these concerns, the SEC adopted new rules in 2023 to further standardise cybersecurity risk disclosures, which includes requirements for reporting on governance and risk management strategies \cite{sec2023cybersecurity}. 

\begin{figure}
    \centering
    \includegraphics[width=1\linewidth]{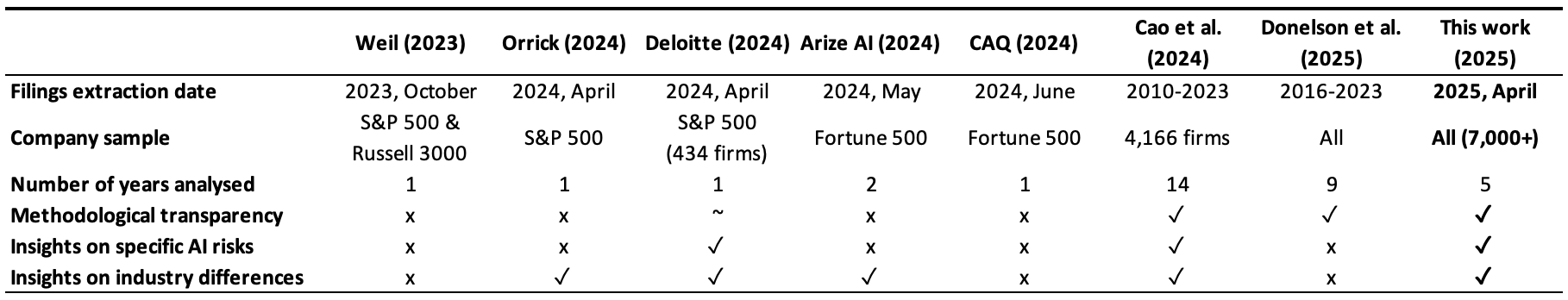}
    \caption{Comparison table of existing reports on AI-mentions in SEC 10-K filings}
    \label{fig:reports2}
\end{figure}

Despite widespread public concerns about AI risks, only a handful of (mostly industry-led) studies empirically explored AI-related mentions in 10-K filings \cite{orrick2024aifilings,weil2023aidisclosures,deloitte2024uscmarshall,arize2024rise,caq2024aireporting,cao2024information,donelson2025strategic}. 
These studies generally find rapid increases in AI (risk) mentions in recent years, with legal, regulatory, competitive and cybersecurity risks emerging as the most common AI risk themes. However, the industry-led reports generally lack methodological rigour and transparency (such as listing the keywords used, or describing how risks are assessed), and often do not go beyond high-level frequency analysis of a single year. Plus, none of the existing studies use data from 2025, or consider (in detail) how societal/ethical AI risks, specifically, are addressed. A summary comparison of existing reports and studies is provided in Figure \ref{fig:reports2}. 

In contrast, our study aims to provide a more comprehensive, systematic, and transparent analysis of AI risk disclosures across five years of 10-K filings. We seek to provide a clearer understanding of how AI risks are currently perceived, framed, and reported in legally mandated corporate disclosures.

% \begin{table}
% \centering
% \noindent\hspace*{-1cm}%
% \resizebox{\dimexpr\textwidth+2cm\relax}{!}{%
% \begin{tabular}{lcccccc}
% \toprule
%  & \textbf{Weil} & \textbf{Orrick} & \textbf{Deloitte} & \textbf{Arize AI} & \textbf{CAQ} & \textbf{Cao et al} & \textbf{Donelson et al} &\textbf{This work} \\
%  & \cite{weil2023aidisclosures} & \cite{orrick2024aifilings} & \cite{deloitte2024uscmarshall} & \cite{arize2024rise} & \cite{caq2024aireporting} & \cite{cao2024information} & \cite{donelson2025strategic} \\
% \midrule
% \textbf{Filings extraction} & 2023, October & 2024, April & 2024, April & 2024, May & 2024, June & ? & ? & \textbf{2025, April}\\
% \textbf{Company sample} & S\&P 500 \& Russell 3000 & S\&P 500 & S\&P 500 (434) & Fortune 500 & Fortune 500 & 4,166 firms & All & \textbf{All (7,000+)} \\
% \textbf{Number of years analysed} & 1 & 1 & 1 & 2 & 1 & 14 & 9 & 5 \\
% \textbf{Methodological transparency} & x & x & $\circ$ & x & x & \checkmark & \checkmark & \textbf{\checkmark} \\
% \textbf{Insights on specific AI risks} & x & x & \checkmark & x & x & \checkmark & x & \textbf{\checkmark} \\
% \textbf{Insights on industry differences} & x & \checkmark & \checkmark & \checkmark & x & \checkmark & x & \textbf{\checkmark} \\
% \bottomrule
% \end{tabular}%
% }
% \caption{Comparison table of existing reports on AI-mentions in SEC 10-K filings}
% \label{fig:reports}
% \end{table}

\section{General trends in AI disclosures}
This section addresses \textbf{RQ1:} \textit{How commonly do public US companies disclose AI use and risks in their 10-K filings, and how has this evolved over the past five years?}

Our aim is to provide a systematic overview of how widespread AI-related disclosures have become in SEC 10-K filings and how the framing of these disclosures -- particularly in terms of AI risks versus AI as part of the companies main business operations -- may have changed in recent years due to technological developments, growing societal concerns and increasing regulatory scrutiny in the field of AI.

\subsection{Methodology: large-scale keyword analysis}
To examine how AI-related disclosures in 10-K filings have evolved over time, we conducted a large-scale keyword analysis of all SEC 10-K filings submitted by publicly traded US companies over the last five years. These filings cover reporting years 2020 through 2024 and amount to an average of approximately 7,000 filings per year. The data was extracted on 1 April 2025, by which point companies following a calendar fiscal year were required to have submitted their 2024 filings. 

The SEC requires that 10-K filings follow a standardised structure, as illustrated in Figure \ref{fig:10k} of Appendix A. Next to covering overall AI-related mentions, we focus our trend analysis in particular on two sections of the 10-K filings: the \textit{Business} section and the \textit{Risk Factors} section, as detailed in Section 2.1. 
These two sections are particularly relevant to understand how companies look at AI developments. Mentions in the Business section likely indicate that firms view AI as a strategic asset or core part of their operations. Mentions in the Risk Factors section, on the other hand, reveal whether companies perceive AI risks to be significant to their business operations or securities \cite{sec2024a10k}. Our methodological choice is also supported by the fact that we found that 91\% of AI mentions occur in these two sections.\\

\noindent\textbf{Data Collection and Tooling}\\
To collect and structure the filings, we extended the existing Edgar-Crawler tool for downloading and parsing SEC filings \cite{loukas-etal-2021-edgar-corpus-and-edgar-crawler}. To this end, we build a custom extension designed specifically for large-scale keyword analyses of SEC 10-K filings. Our extension supports structured keyword extraction using regular expressions, sentence-level filtering and output formatting in an Excel format to facilitate easy inspection and annotation. 
To improve accessibility of SEC-filings, we also developed a web-based interface for our tool that enables non-technical users to search SEC filings by keywords, companies, year, and section, and export matched results directly in Excel format (see Figure \ref{fig:tool}) in Appendix G). The \href{https://github.com/lucas-ubm/sec_analysis}{tool} -- including code and documentation -- is publicly available to ensure reproducibility and support future research.\\

\noindent\textbf{Keyword matching}\\
Using this tool, we applied a list of AI-related keywords (e.g., AI, artificial intelligence, machine learning, deep learning) to the full dataset. The complete keyword list is added in Appendix A. This process identified over 110,000 keyword matches and 50,700 unique AI-related sentences across all filings. 

To assess the precision of our keyword matching, we manually reviewed a random sample of 385 keyword matches. Only two were found not to refer to artificial intelligence. Based on this, we estimate with 95\% confidence that at least 98.78\% of our 110,000 keywords matches are truly AI-related. As this evaluation method does not estimate recall, our findings likely represent a slightly conservative estimate of how often companies refer to AI in their disclosures. 

\subsection{Results: How are companies considering AI in their 10-K filings?}
We assess how companies disclose AI in their 10-K filings using two main indicators: the \textit{percentage of companies} mentioning AI at least once (Table \ref{tab:mentions}), and the \textit{average number of unique AI-related sentences per filing} among those that include at least one AI-mention (Table \ref{tab:mentions}). These metrics capture both the spread and prominence of AI in corporate disclosures. Both metrics are normalised by the number of filings per year, to account for the changing number of companies each year.

The share of companies reporting on AI in their filings has more than doubled between 2022 and 2024, with over 50\% of companies now mentioning AI, indicating that AI is becoming a more standard and important topic for companies. At the same time, the average number of AI-related sentences per 10-K filing (with at least one AI-mention) has steadily increased each year, showing - next to a rise in companies talking about AI - also the growing prominence of the topic in disclosures.\\

\begin{table}[t]
\centering
\begin{tabular}{l r r}
\toprule
\multirow{1}{*}{\textbf{Reporting}} & 
\textbf{Companies} & 
\textbf{Avg. unique} \\
\textbf{year} & \textbf{that mention AI} & \textbf{sentences about AI} \\
\midrule
2020 & 12.97\% & 4.96 \\
2021 & 17.10\% & 5.98 \\
2022 & 18.89\% & 6.25 \\
2023 & 34.90\% & 7.48 \\
2024 & 50.60\% & 9.27 \\
\bottomrule
\end{tabular}
\vspace{0.5em}
\caption{Evolution of mentions of AI in 10-K forms over time. The table shows a clear trend towards more AI mentions over time.}
\label{tab:mentions}
\end{table}

\noindent\textbf{Risk vs business mentions}\\
We next analyse how companies frame AI in their filings by comparing AI-mentions across the Business and Risk Factors sections. 

As shown in Figure \ref{fig:risks_business}, mentions of AI in the Business section have increased steadily: from 12\% of firms in 2020 to nearly 25\% in 2024. This indicates AI's growing role in company's operations, especially considering the recent warnings and enforcement actions by the SEC regarding ``AI washing'' in disclosure filings.  

More notably, AI risk disclosures have grown more dramatically, confirming trends from existing reports over the previous years. In 2024,  43\% of companies mentioned AI in their Risk Factors section --  a sevenfold increase since 2022. At the same time, the total number of AI risk-related mentions has grown by more than 950\% from 2022 to 2024. We see an average of 2.7 unique AI-risk sentences per company in 2024. 
These trends suggest that firms are not only adopting AI in their business operations, but are also increasingly aware of the risks posed by AI. 

Interestingly, 26\% of all companies mention AI in their Risk Factors section but not in their Business section. This suggests that even firms that likely do not view AI as a core part of their business operations may still perceive significant risks associated with it. By contrast, only 7\% of companies mention AI in their Business section without addressing it in their Risk section.\\ 

\begin{figure}[h]
    \centering
    \includegraphics[width=0.95\linewidth]{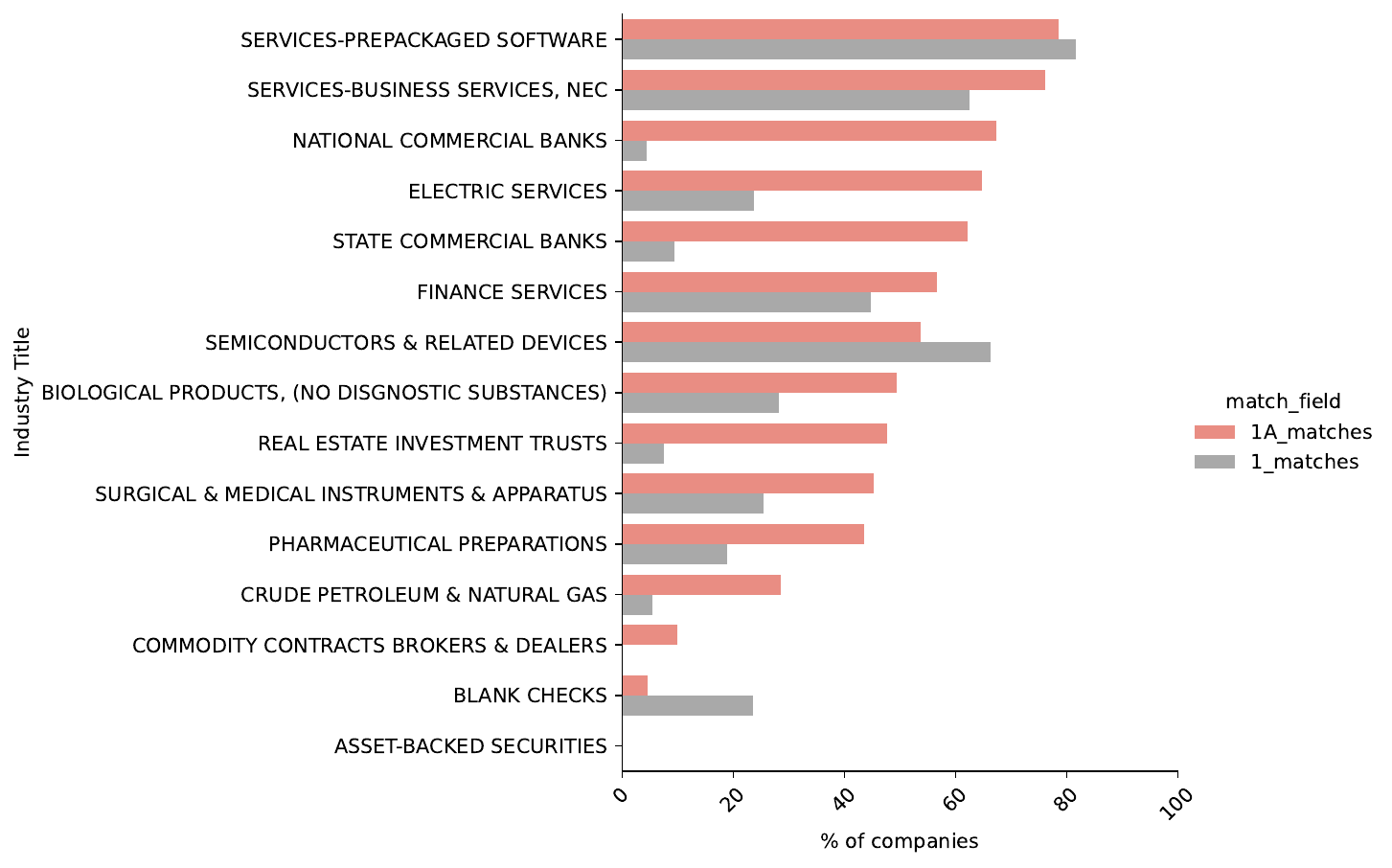}
    \caption{Percentage of companies mentioning AI in Risks vs. Business sections of 10-K filings in 2024 for the top 15 most represented industry sectors (per SIC code) in 10-K filings %\konrad{i could not read this and enlarged it.}
    }
    \label{fig:top15}
\end{figure}

\noindent\textbf{Differences across industries}\\
Using SEC industry classification codes (SIC), we analyse how AI disclosures vary across sectors. Figure \ref{fig:top15} presents the percentage of companies within the 15 most represented industries (based on 10-K filings) that mention AI in their Business and Risk sections.
As expected, software companies (e.g., Apple, SAP, Cisco, Microsoft) lead in both categories, with over 75\% referencing AI in both sections. High percentages of AI disclosure in the business section are also found in the business services, finance services, pharmaceutical, semiconductor, and medical sectors. In contrast, industries such as banking, real estate, and electrical equipment have relatively low levels of AI-related business mentions, despite more frequent acknowledgments of AI-related risks.

Looking at trends over time, we observe sharp increases in AI-related risk disclosures between 2022 and 2024 in the pharmaceutical, electric services, banking, and real estate sectors. Given the size of these sectors, this contributes significantly to the overall growth in the percentage of companies disclosing AI risks. Detailed industry trends over time are presented in Figure \ref{fig:sectoral_risk_trends} of Appendix D.
\section{Qualitative analysis of AI risk disclosures}
Building on the trend analysis of Section 3, we now turn to examining the content of AI risk disclosures. To address \textbf{RQ2}: \textit{What types of AI-related risks do public US companies disclose and how do these disclosures align with broader societal concerns about AI?}, we conduct a qualitative analysis of a subset of AI-related risk statements.

\subsection{Methodology}
From the AI-related risk sentences extracted in Section 3, we selected a sample of statements for the manual qualitative analysis. The sample includes filings from companies across three different groups: 
\begin{enumerate}
    \item\textbf{Random Group:} A random sample of 20 companies from various industries and sizes, to capture general patterns in AI risk narratives. 
    \item\textbf{Top-Tech Group:} The 10 largest publicly listed US tech firms (by market capitalisation), to examine how leading tech and AI companies frame AI risks and whether these align with societal concerns. 
    \item\textbf{Risk Only Group:} A random sample of 20 companies that mentioned AI in their Risk section for the first time in 2024, while not mentioning AI in their Business section. This group provides insight into disclosures from firms that likely do not view AI as core part of their operations. 
\end{enumerate}
A full list of the 50 companies is provided in Appendix B.

We focused our qualitative analysis on the most recent filings (reporting year 2024), given the rising prominence of AI risks and the lack of existing studies that cover 2024 data. Our sample included 495 AI-related sentences: 42\% from the Top-Tech Group, 37\% from the Random Group, and 22\% from the Risk Only Group. After removing non-risk-related content -- such as section headers or vague mentions like “we use AI” -- 444  sentences remained for the qualitative analysis.\\ 

\noindent\textbf{Qualitative analysis of the data}\\
Next, we developed a structured labelling framework that combines a bottom-up data-driven approach with a top-down, theory-informed taxonomy, which allowed us to systematically analyse both business-centric and societal AI risk statements. 

We began with an inductive thematic analysis \cite{clarke2014thematic} of a sample of AI-related risk statements (drawn from all three groups). Two researchers independently reviewed and coded this sample, using the categories from the existing reports on AI risks in 10-K filings as inspiration. We then iteratively developed a shared set of categories through discussion and refinement. This process yielded four dominant themes, three of which reflect business-orientated risks: (1) legal risks, (2) competitive risks, and (3) reputational risks. The other statements were grouped under the broader term of societal-relevant AI risks. Using the same strategy, we further developed sub-codes to capture frequently recurring topics for the business-oriented risks (e.g., regulatory uncertainty, IP issues, high investment costs).

To systematically analyse the disclosures of the societal-relevant AI risk theme, we complemented our inductive bottom-up approach with a theory-driven approach. Specifically, we used the MIT AI Risk Domain Taxonomy developed by Slattery et al. \cite{slattery2024ai}, which offers a comprehensive set of AI risk domains and subcategories. This Taxonomy was developed on the basis of a synthesis exercise of over 60 AI risk frameworks and a large database of AI risks and incidents. While this taxonomy is not specifically designed to cover societal AI risks, many domains (e.g., misinformation, malicious use, technical limitations) align closely with societal concerns around AI, especially when seen in the light of the global-scale of operations of the companies in our samples. We therefore use this taxonomy to assess which societal risks are disclosed. 

An overview of our four top-level risk category labels -- legal, competitive, reputation and societal -- is shown in Table \ref{tab:ai_risk_categories}, along with the sub-labels used in the coding process. To allow for a more in-depth assessment of the societal AI risk statements, we additionally used the subdomain categories of the MIT AI Risk Taxonomy in our labelling process, an overview of these are shown in Figure \ref{fig:mit_taxonomy} Appendix E.\\

\begin{table}
\centering
\begin{tabular}{>{\centering\arraybackslash}p{3.5cm} p{6cm}}
\toprule
\textbf{General risk categories} & \textbf{Subcategories} \\
\midrule
\multirow{5}{3.5cm}{\textbf{Legal AI risk}} & Compliance costs \\
& Legal uncertainty \& complexity \\
& (Potential) legal actions \& liability \\
& IP concerns \\
& Other / general \\
\midrule
\multirow{3}{3.5cm}{\textbf{Competitive AI risk}} & Rapid developments \\
& Large investments needed \\
& Other / general \\
\midrule
\textbf{Reputational AI risk} & - \\
\midrule
\multirow{7}{3.5cm}{\textbf{Societal AI risk}} & Discrimination \& toxicity \\
& Privacy \& security \\
& Misinformation \\
& Malicious actors \& misuse \\
& Human-computer interaction \\
& Socioeconomic \& environmental harms \\
& AI system safety, failures \& limitations \\
\bottomrule
\end{tabular}
\caption{An overview of the main categories and subcategories that were used in our labelling process, derived from the systematic inductive analysis of the data.}
\label{tab:ai_risk_categories}
\end{table}

\noindent\textbf{Labelling process}\\
Each sentence was annotated with one or more of the primary and sub-labels based on its content, hereby taking into account adjacent sentences and the broader context if necessary. The annotation was performed independently by two authors, using the shared coding guide. The inter-annotator agreement for the overall categories was: 90\% for legal risks, 82\% for competitive risks, 95.5\% for reputational risks and 93\% for societal risks. To account for agreement due to random chance, we also report Cohen's $\kappa$ \cite{cohen_coefficient_1960} for the risk categories, we obtain: 0.80 for the legal risks, 0.60 for competitive risks, 0.75 for reputational risks and 0.82 for societal risks. Disagreements were resolved through joint discussion. The results are presented in the next section. 

\subsection{Results: What AI risks do companies disclose?}
\begin{figure}
     \centering
     \includegraphics[width=0.75\linewidth]{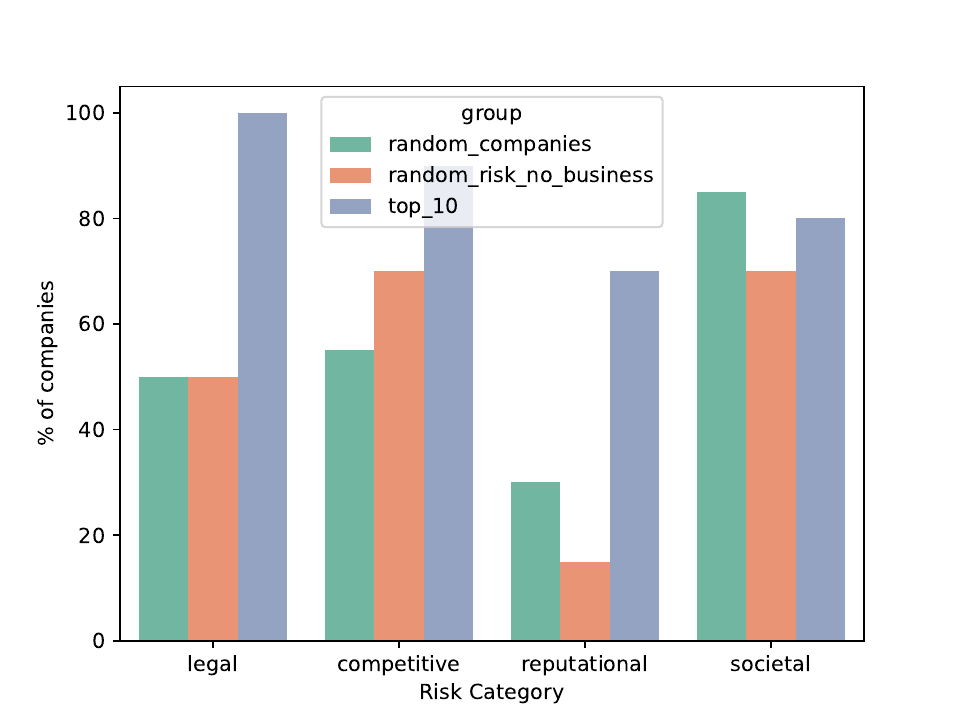}
     \caption{The percentage of companies in each sample group that mention legal, competitive, reputational, or societal AI risks in their 10-K filings.}
     \label{fig:labels_highlevel}
 \end{figure}

\noindent\textbf{General overview of disclosed risk types}\\
Figure \ref{fig:labels_highlevel} shows how often companies in each group mention legal, competitive, reputational, or societal AI risks. Most top tech companies disclose all four risk types, though many firms in the other two groups also cite multiple categories, including societal AI risks. 

Table~\ref{fig:average_sentences} reports the average number of AI-related sentences per risk category across the three company groups. Legal and competitive risks are the most commonly discussed, particularly among top tech firms. 
\begin{table}
    \centering
    \includegraphics[width=0.9\linewidth]{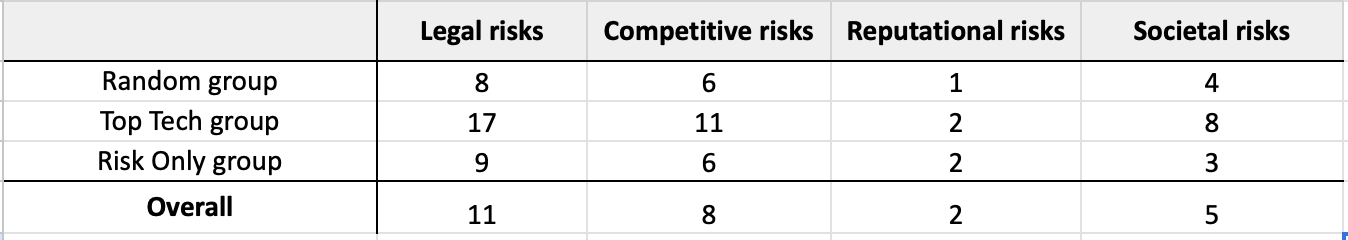}
    \caption{The average number of sentences for each risk category per filing}
    \label{fig:average_sentences}
\end{table}

\noindent\textbf{Legal and competitive risks}\\
Legal AI risks are the most widely disclosed. These cover regulatory uncertainty, increasing regulatory complexity, compliance burdens, and liability and IP issues. Legal AI risks are most frequently mentioned by the large tech firms, but the results show that such risks are generally shared broadly across all publicly traded firms. An example is the asset management company Blackstone Infrastructure Strategies, stating: ``\textit{the EU adopted the Artificial Intelligence Act in 2024, ... which may create additional compliance burdens, higher administrative costs and significant penalties}''.

Competitive AI risks are also widely shared. Many firms, especially in the top-tech sample, discuss the large uncertain investments that are needed to compete or stay ahead in the AI race. Firms in the random samples similarly express competitive AI concerns, including falling behind in AI innovation, losing market share due to new AI competitors, or failing to adopt AI fast enough \\

\noindent\textbf{Societal AI risks}\\
Companies mention a wide range of societal AI risk, of which the most common themes include misuse by others (e.g., cyberattacks or fraud) and internal privacy/security issues due to AI use. One example is Baker Hughes, an energy technology company, stating ``\textit{Cyberattacks are expected to accelerate on a global basis in both frequency and magnitude as threat actors become increasingly sophisticated in techniques and tools (including artificial intelligence)}''. Figure \ref{fig:subdomains} in Appendix F presents the full breakdown of mentions per societal risk domain and sub-domain. 

Societal risk mentions are not exclusive to top tech firms. For example, technical limitations of AI systems (opacity, bias, robustness) are more frequently disclosed by non-tech companies. Similarly, risks related to discrimination and content toxicity are mentioned by about 30\% of companies in both the random and top tech samples. In contrast, misinformation risk mentions appear more frequently in the top-tech group, likely due to the social media companies that are included in this sample.

Some societal risk types remain absent in the disclosure data. We found no disclosures related to the environmental harms of AI, socioeconomic displacement due to AI, dangerous AI capabilities, multi-agent risks, or the pollution of the information ecosystem (e.g., via generative AI content). Figure~\ref{fig:subdomains} in Appendix F provides a full overview of disclosure statistics for each subdomain.\\

\noindent\textbf{The framing of AI risk disclosures}\\
We also observed clear patterns in how risks are framed. The top-tech firms often seem to externalise societal AI risks, attributing them to third-party misuse (e.g., faulty datasets or misuse of their models), while rarely acknowledging their own role in developing and deploying systems that may contribute to these risks, particularly those stemming from technical limitations (e.g., bias, accuracy). When such risks are mentioned, they are typically described in general, or hypothetical terms.
%, with little discussion of mitigation strategies. 
For example, Meta notes: ``\textit{There are significant risks involved in developing and deploying AI and there can be no assurance that the usage of AI will enhance our products or services or be beneficial to our business, including our efficiency or profitability}'' and ``\textit{We may not have insight into, or control over, the practices of third parties who may utilize such AI technologies}''.

The reliance on broad boilerplate language is common across all groups. However, we also find examples of companies that do offer more concrete acknowledgments of AI’s technical limitations and related societal risks, including some details on risk mitigation strategies. 
For instance, Blackstone Infrastructure Strategies states: ``\textit{Moreover, with the use of AI Technologies, there often exists a lack of transparency of how inputs are converted to outputs, and neither the Sponsor nor any Portfolio Entity can fully validate this process and its accuracy}''. Similarly, Cognizant Technology Solutions notes: ``\textit{The uncertainty around the safety and security of new and emerging AI applications requires significant investment to test for security, accuracy, bias, and other variables - efforts that can be complex, costly, and potentially impact our profit margins}''.
\section{Discussion}
Our findings show that \textbf{companies are increasingly disclosing AI-related risks} in their 10-K filings (RQ1). Between 2022 and 2024, the share of firms grew sevenfold, with \textbf{over 43\% of firms now mentioning AI risks}. At the same time, the number of AI-related sentences per 10-K filing (with at least one AI-mention) has also steadily increased, showing the overall increasing prominence of AI in corporate disclosures.
Interestingly, more companies now describe AI as a material risk (43\% of companies) than as part of their main business operations (only 23\% of companies). This disparity may reflect the growing regulatory scrutiny by the SEC on misleading ``AI washing'' statements in disclosures.

\textbf{In terms of substance (RQ2), legal and competitive risks dominate corporate AI risk disclosures} (confirming existing reports from previous years). Many firms -- including those not directly involved in AI development -- cite regulatory uncertainty, compliance burdens, and liability concerns. Competitive risk disclosures mainly focus on fears of falling behind in the AI race, particularly among top tech firms that highlight the large investments needed.

\textbf{We also found a wide range of societal risks being discussed across companies} -- most notably malicious uses of AI (e.g., cyberattacks or fraud), internal privacy and security breaches, and technical limitations of AI systems (e.g., reliability and transparency). However, several key societal AI issues, such as AI’s environmental impact, labour market effects, dangerous capabilities, and the pollution of the information ecosystem, are largely absent in the data. 

\textbf{Moreover, AI risks are often framed in vague, hypothetical, or boilerplate language}, 
Furthermore, many tech companies seem to externalise societal AI risks, for example attributing them to third-party misuse, while rarely acknowledging their own role in developing the technologies that may contribute to those risks. These patterns mirror long-standing trends observed in prior research on cybersecurity disclosures, which consistently find firms relying on generic and boilerplate language to describe cyber risks.

These findings have several implications. First, they highlight the limitations of 10-K filings as a source for understanding companies' stance on AI risks. As firms are only required to disclose material risks to their business or investors, broader (societal relevant) AI harms may remain undiscussed. Legal liability concerns may further discourage firms from acknowledging technical limitations or controversial impacts of AI tools.

Second, our findings indicate that the SEC’s recent push for more truthful and tailored AI disclosures seems clearly warranted. More specific rules and guidance -- similar to those introduced in cybersecurity -- could help improve the quality of risk disclosures. This includes encouraging firms to detail risk mitigation strategies, especially for companies central to the development of high-impact AI systems.

\textbf{Limitations.} This study focuses exclusively on US' SEC data, which may limit the generalisability of our findings. Our qualitative analysis is based on a relatively small subset of 50 companies, which may not capture the full range of disclosures. In addition, our keyword-based extraction method may have overlooked other relevant disclosure statements, or missed important contextual nuances. Lastly, we do not use a systematic method for quantifying the prevalence of boilerplate language in AI risk disclosures. Future work could address this gap by developing metrics to assess disclosure specificity, and comparing the results with more established domains, such as cybersecurity. 

\section{Conclusion}
Our large-scale analysis of over 30,000 SEC 10-K filings reveals a sharp increase in AI-related risk disclosures in recent years, with over 43\% of firms now mentioning AI risks. Legal and competitive concerns are most frequently cited, but we also observe increasing attention to societal AI risks, including frequent mentions of AI-enabled cyberattacks, fraud, bias, and technical limitations of AI systems. However, many disclosures remain vague and rely on general, boilerplate language. These patterns reinforce recent concerns raised by the SEC about the quality and specificity of AI disclosures. Therefore, the SEC should consider introducing more targeted disclosure rules and guidance, similar to those in cybersecurity, particularly for firms more directly involved in AI development.

Future research should explore the quality and specificity of AI risk disclosures in greater detail. Comparative analyses with EU data could also prove insightful, especially with recent EU regulations that have introduced AI risk assessment and reporting requirements. 
To support such future research, we introduce an open-source web tool that allows for easily extracting and analysing topical disclosures across SEC filings. 
%\konrad{i move future resesarch here because this made more sense.}
\section*{Acknowledgements}
The authors are supported by the RegTech4AI AiNed Fellowship Grant, which is funded by the Dutch National Growth Fund (NGF) under file number NGF.1607.22.028.

We are thankful to our colleagues at the Maastricht Law \& Tech Lab: Gijs van Dijck, Kamil Szostak, and Tom Vos, for their support in setting-up the data infrastructure, refining the research objectives and providing feedback throughout the project.  

We also acknowledge the use of AI tools, including ChatGPT, Gemini and Visual Studio Code extensions, which supported various aspects of the code development and text editing.


\bibliographystyle{splncs04}
\bibliography{references.bib}
\newpage
\subsection*{Appendix A: An example of a 10-K filing}
\begin{figure}[h!]
    \centering
    \includegraphics[width=0.95\linewidth]{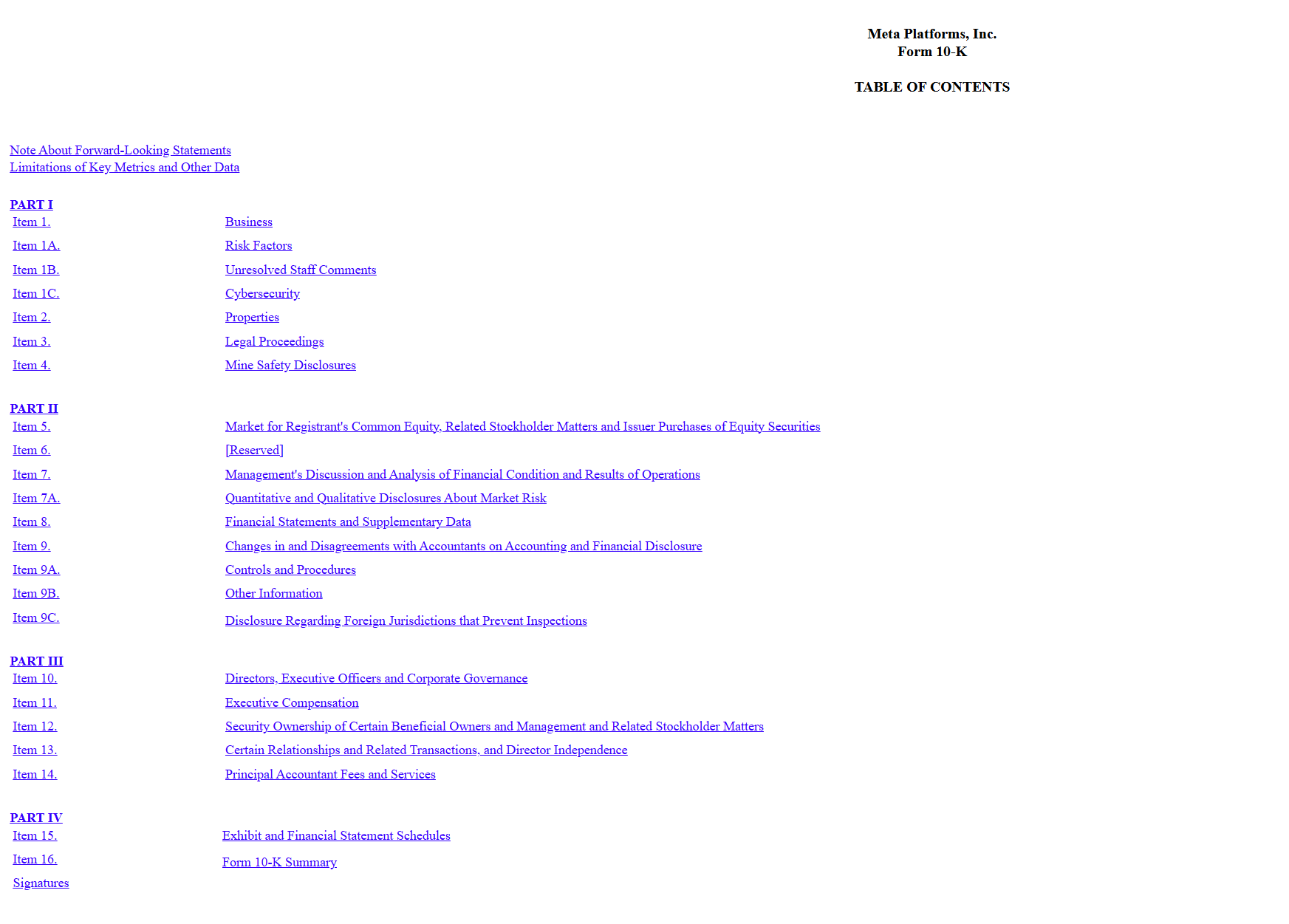}
    \caption{Example of the start of a 10-K filing (Meta, 2024), illustrating the standard structure mandated by the SEC.}
    \label{fig:10k}
\end{figure}

\subsection*{Appendix B: The AI-related keywords used to extract sentences}
The complete list of keywords used to extract all AI-related sentences from the dataset (using regular expressions).
\begin{table}[h!]
\centering
\begin{tabular}{|l|l|l|}
\hline
Artificial Intelligence & Deepfake                 & Natural Language Processing \\
\hline
Machine Learning        & Deep Learning            & Computer Vision \\
\hline
Generative              & Image Recognition        & Speech Recognition \\
\hline
Voice Assistant         & Chatbot                  & Recommendation System \\
\hline 
A.\*I & NLP & Recommender System  \\
\hline 
Artificial General Intelligence & AGI & \\
\hline
\end{tabular}
\label{tab:ai_keywords}
\end{table}
\subsection*{Appendix C: Sample of companies included in annotation analysis}
The top 10 tech companies included were: Amazon Com Inc, Nvidia Corp, Netflix Inc, Tesla Inc, Meta Platforms Inc, Oracle Corp, Alphabet Inc, Broadcom Inc, Apple Inc, Microsoft Corp. 

The random 20 companies were: Cognizant Technology Solutions Corp, Constellation Energy Generation LLC, CCO Holdings Capital Corp, Graham Alternative Investment Fund II LLC, Live Oak Bancshares Inc, Pinterest Inc, Premier Inc, Avidity Biosciences Inc, Invitation Homes Inc, Rekor Systems Inc, Baker Hughes Co, Playtika Holding Corp, Venture Global Inc, Blackstone Infrastructure Strategies LP, American Airlines Inc, Micron Technology Inc, Webster Financial Corp, Royal Gold Inc, Mind Technology Inc.

And the random 20 companies with no business mentions were: Sealed Air Corp/DE, Biomarin Pharmaceutical Inc, Eagle Bancorp Inc, Nlight Inc, Viper Energy Inc, Zentalis Pharmaceuticals Inc, Compass Therapeutics Inc, Third Coast Bancshares Inc, Bitcoin Depot Inc, LGAM Private Credit LLC, New Era Helium Inc, Textron Inc, UDR Inc, Healthpeak Properties Inc, Newell Brands Inc, Financial Institutions Inc, Autozone Inc, Credit Acceptance Corp, Kohls Corp, Helen of Troy Ltd. 

\subsection*{Appendix D: Industry-level trends in AI risk disclosures}
To explore how AI risk mentions have evolved across sectors, we analysed changes in the proportion of companies reporting AI risks across the 15 most common industry sectors (based on SEC SIC codes). 
\begin{figure}[h!]
\centering
\includegraphics[width=0.99\linewidth]{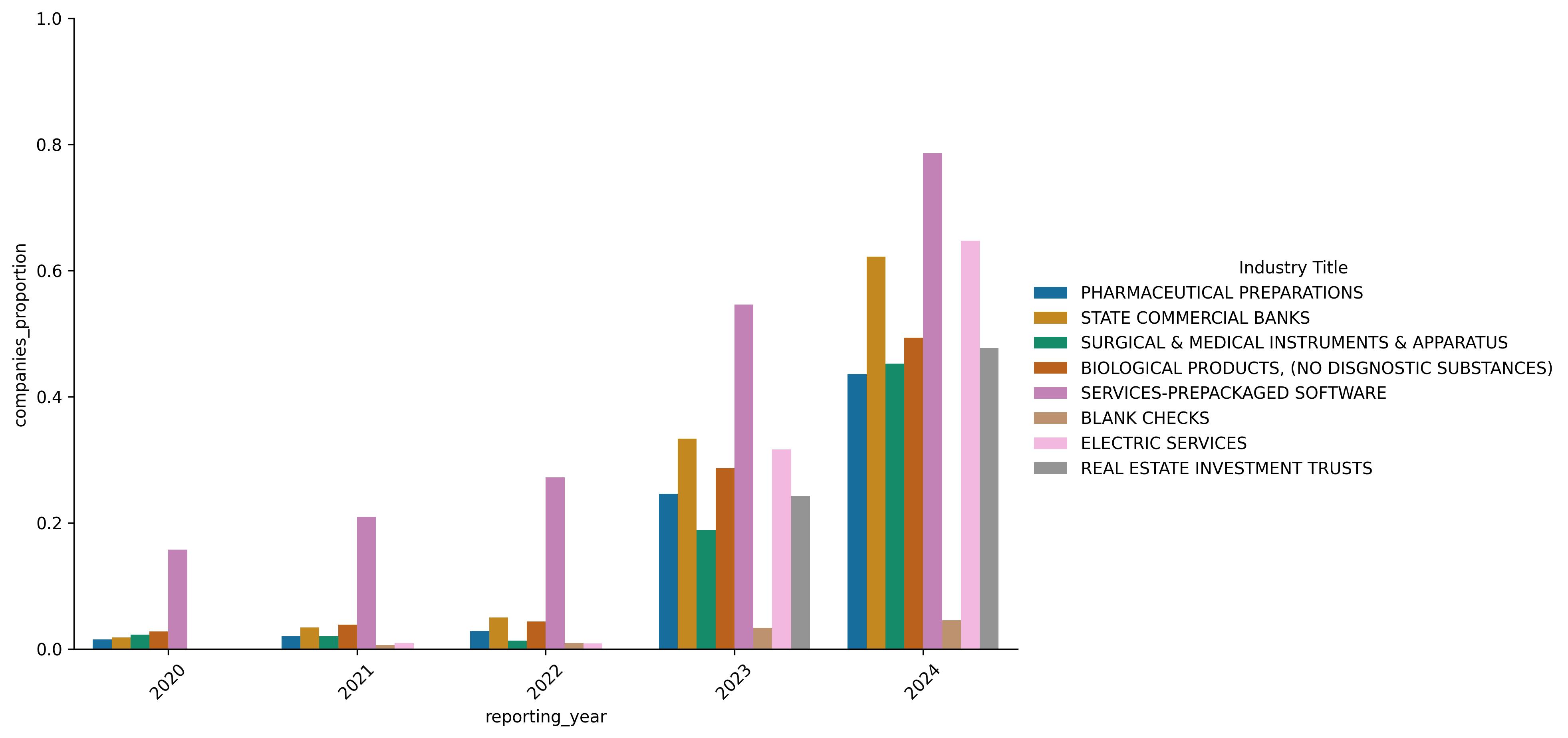}
\caption{Change in the proportion of companies reporting AI risks over time, by industry sector (2020–2024). Sector size is not normalized.}
\label{fig:sectoral_risk_trends}
\end{figure}

\subsection*{Appendix E: Annotation coding framework}
\begin{figure}[h!]
    \centering
    \includegraphics[width=0.8\textwidth]{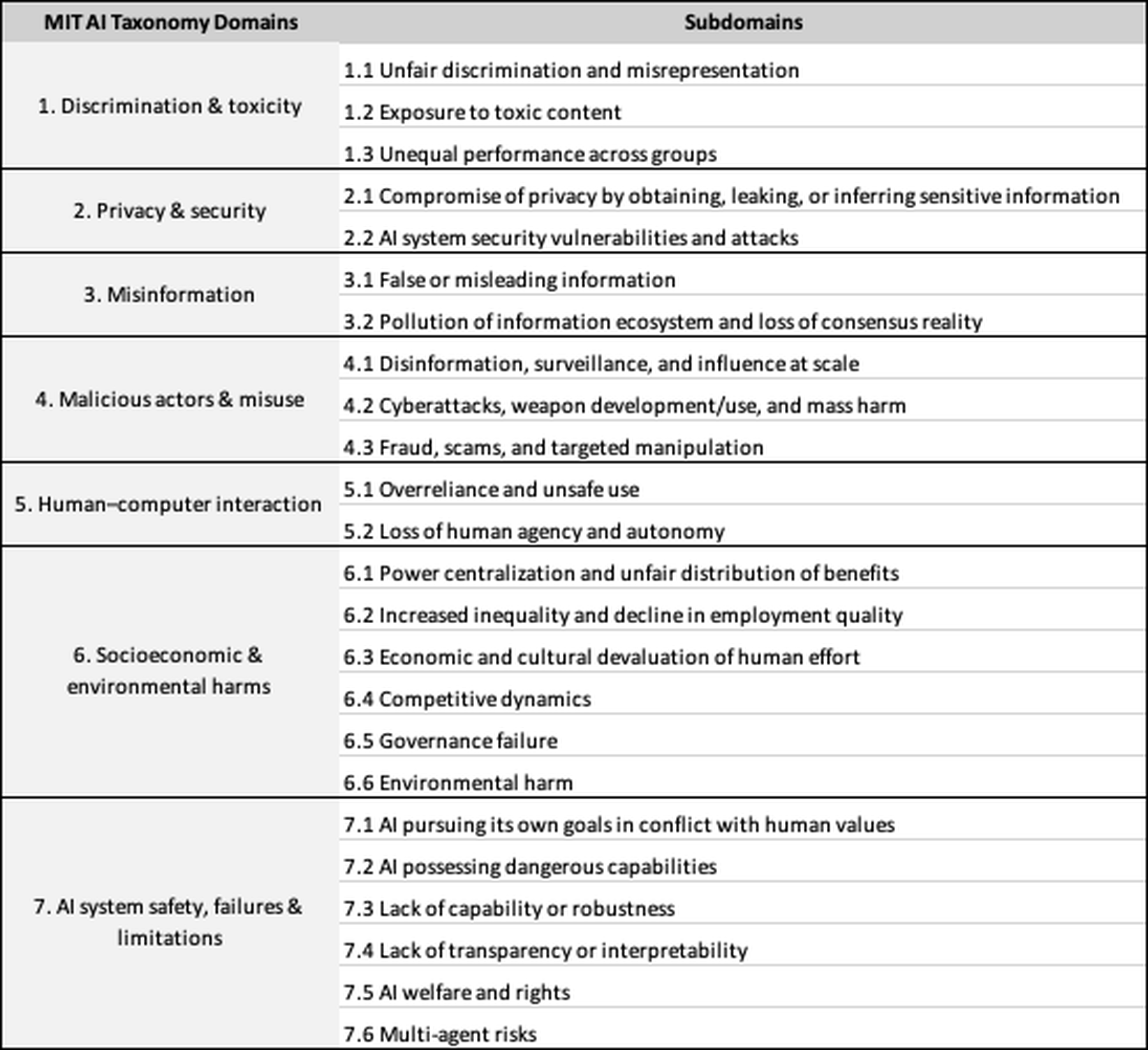}
         \caption{MIT AI Risk Taxonomy including subdomains (adapted from \cite{slattery2024ai})}
         \label{fig:mit_taxonomy}
\end{figure}

We used a structured labelling scheme to manually classify the content of AI risk disclosures. The framework combines general risk categories (legal, competitive, reputational, societal) with fine-grained subcategories, including those from the MIT AI Risk Taxonomy (see also Table \ref{fig:mit_taxonomy}. Table \ref{fig:mit_taxonomy} below provides a detailed view of the MIT AI Risk Taxonomy domains and subdomains used in the annotation of societal risk statement \cite{slattery2024ai}. 

\subsection*{Appendix F: Distribution of societal AI Risks by domain subdomain of the MIT AI Risk Domain Taxonomy}
This figure presents the frequency of disclosures across the MIT taxonomy subdomains for the 50 companies in our sample. These include risks related to misinformation, bias, system vulnerabilities, and more. 

\begin{figure}[h!]
\centering
\includegraphics[width=0.9\linewidth]{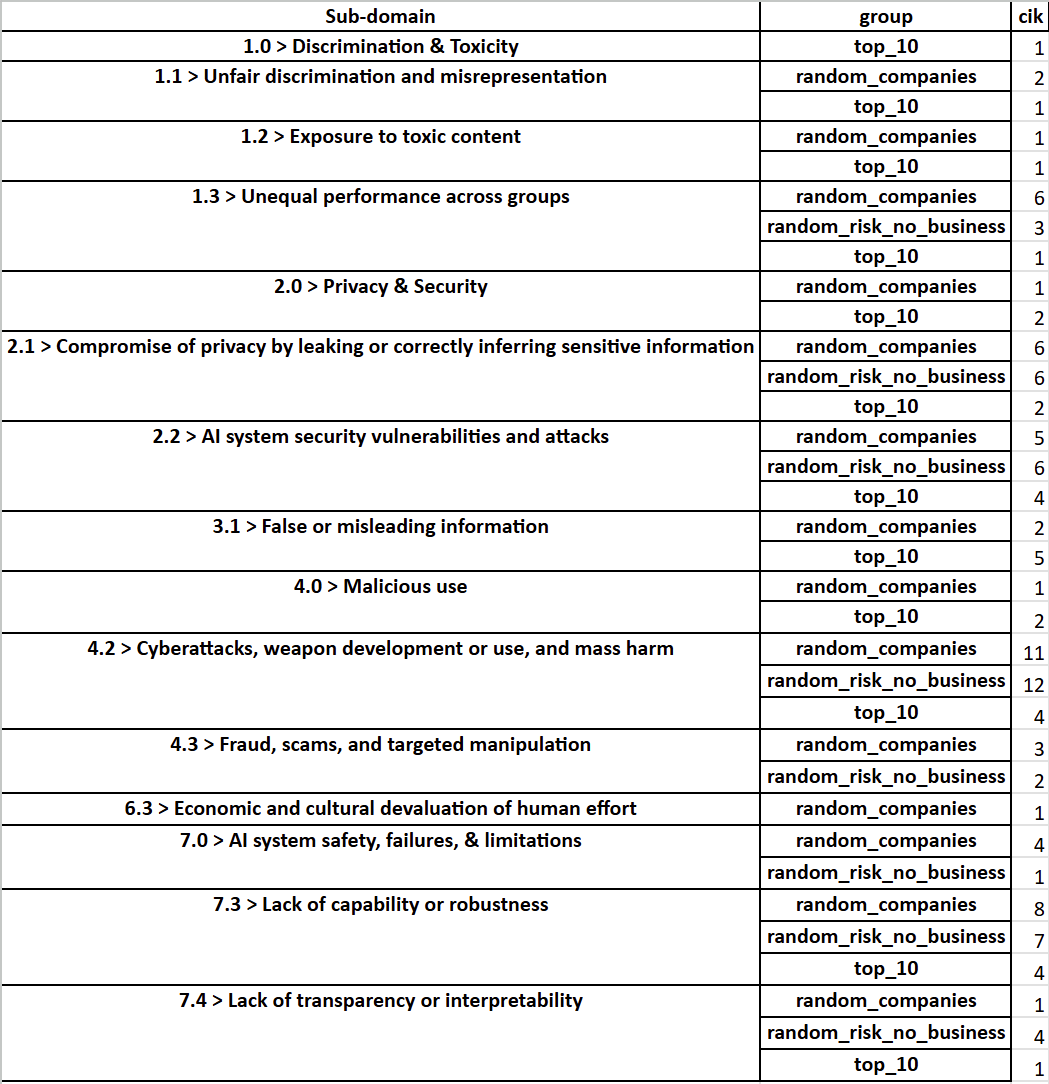}
\caption{Distribution of societal AI risk disclosures across MIT AI Risk Taxonomy subdomains.}
\label{fig:subdomains}
\end{figure}

\subsection*{Appendix G: Interface of the web tool developed in this study}

\begin{figure}[h!]
\centering
\includegraphics[width=0.9\linewidth]{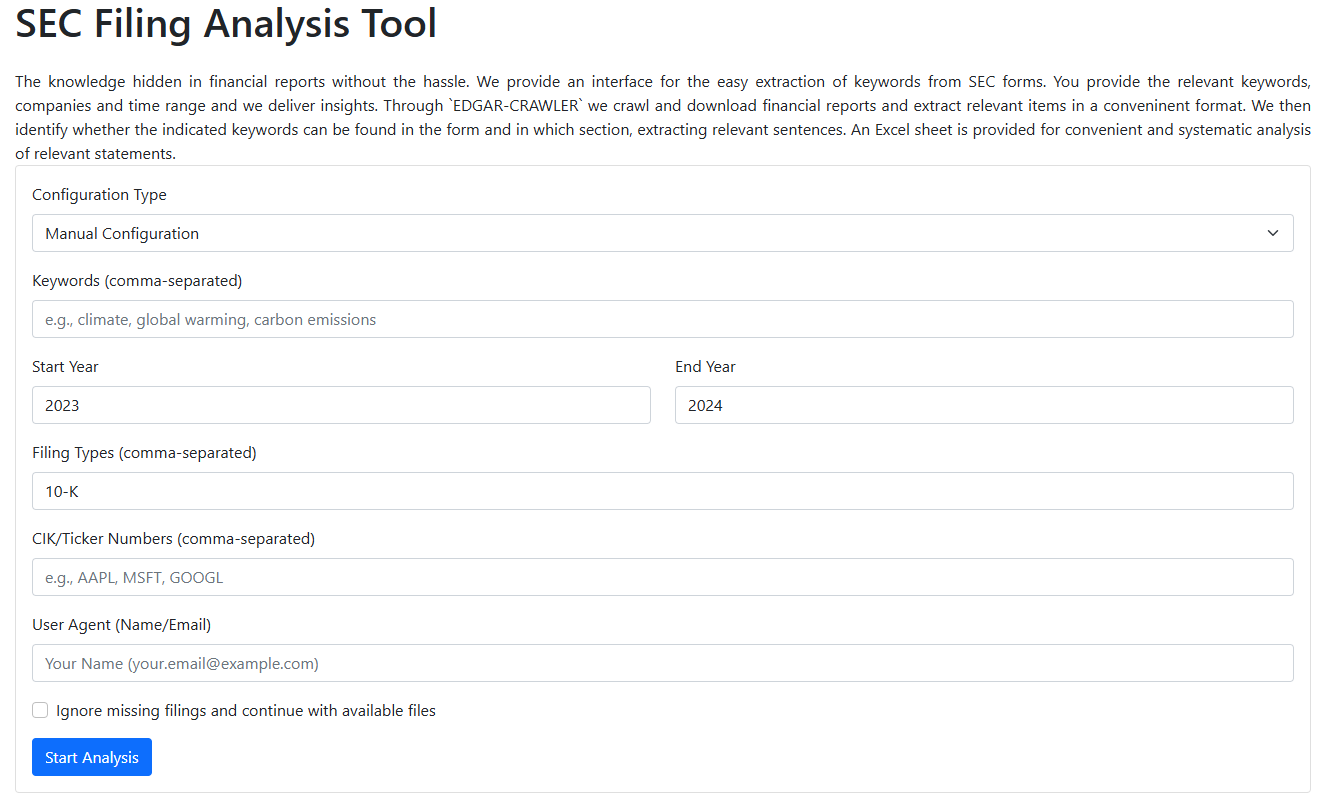}
\caption{Web tool interface}
\label{fig:tool}
\end{figure} % how to not have this numbered? 

\end{document}